\documentclass[nofootinbib,prd,twocolumn,showpacs,showkeys,preprintnumbers]{revtex4-1}
\usepackage{hyperref,amssymb,amsmath,mathrsfs,bm,graphicx}
\begin{document}
\title {Cracking and complexity of self--gravitating dissipative compact objects}
\author{L. Herrera}
\email{lherrera@usal.es}
\affiliation{Instituto Universitario de F\'isica
Fundamental y Matem\'aticas, Universidad de Salamanca, Salamanca 37007, Spain.}
\author{A. Di Prisco}
\email{alicia.diprisco@ucv.ve}
\affiliation{Escuela de F\'{\i}sica, Facultad de Ciencias,
Universidad Central de Venezuela, Caracas 1050, Venezuela.}

\date{\today}
\begin{abstract}
 The concept of  cracking refers to the tendency of a fluid distribution to ``split'', once it abandons the equilibrium. In this manuscript we develop a general formalism to describe the occurrence of cracking within a dissipative fluid distribution, in comoving coordinates. The role of dissipative processes in the occurrence of cracking is brought out. Next, we relate the occurrence of cracking with the concept of complexity for self--gravitating objects defined in \cite{ps1, ps2, epjc}. More specifically we relate the occurrence of cracking   with the condition of the vanishing of the scalar function intended to measure the complexity of the fluid distribution (the complexity factor). We also relate the occurrence of cracking with the specific mode of leaving the equilibrium. Thus, we prove that leaving  the equilibrium  in either, the homologous ($H$), or the quasi--homologous regime ($QH$), prevents the occurrence of cracking. Also, it is shown that imposing the condition of vanishing  complexity factor  alone, (independently of the mode of leaving the equilibrium) prevents the occurrence of cracking  in the non--dissipative geodesic case, and in the non--dissipative isotropic case. These  results  bring out further the relevance of the complexity factor and its related definition of complexity, in the study of self--gravitating systems.

\end{abstract}
\date{\today}
\maketitle

\section{Introduction}
 This work deals with the interplay between  the concepts of cracking  \cite{herrera}  and  complexity \cite{ps1, ps2, epjc}, of a  fluid distribution in presence of dissipative processes.
These  two concepts have been shown to be relevant in the study of self--gravitating systems. 
Besides,  dissipative processes are expected to be present during many phases of the stellar evolution.

The concept of  cracking is associated  to the tendency of a fluid distribution to ``split'', once it abandons the equilibrium as consequence of perturbations. Thus we say that, once the system has abandoned the equilibrium,  there is  a cracking, whenever its inner part tends to collapse whereas its outer part tends to expand. The cracking takes place at the surface separating the two regions. When the inner part tends to expand and the outer one tends to collapse we say that there is an overturning.

In order to avoid some misunderstandings often found in the literature, we find it useful to stress the following remarks: 
\begin{enumerate}
\item The concepts of stability and cracking are  different, although they are often confused.
\item The term stability refers to the capacity of a given fluid distribution to return to  equilibrium once it has been removed from it. The fact that the speeds of sound are not superluminal does not assure in any way the stability of the  object, it only ensures causality.

\item  The cracking only implies the tendency of the system to``split''  immediately after leaving the equilibrium, where ``immediately'' means on a timescale smaller than the hydrostatic timescale, and the thermal relaxation time.  Whatever happens next, whether the system enters into a dynamic regime, or returns to equilibrium, is independent of the concept of cracking. Of course the occurrence of cracking will affect the future of the fluid configuration in either case.

\item In order to check the occurrence (or not) of cracking one must take  the system out of its state of equilibrium. For doing that one submits the system to fluctuations. In the original paper of cracking \cite{herrera} these fluctuations were assumed to be generic (of an``unspecified'' nature). The specific case of fluctuations associated to compression of the fluid has been considered in \cite{abreu}. In this latter case, the confusion between cracking and stability may appear due to the fact that the adiabatic index is related to the speed of sound and the stability. 

\end{enumerate}
In other words:
\begin{itemize}
\item In a system which is stable, i.e. a system  that once removed from equilibrium  comes back to it in a timescale of the order of hydrostatic time, cracking may occur or not.

or 
\item  After the occurrence of cracking the system may return to equilibrium (the system is stable) or enters into a dynamic regime (the system is unstable).
\end{itemize}

It is worth mentioning, with respect to the  physical  relevance of cracking, that cracking might be invoked as the  possible origin of quakes in neutron stars \cite{q1,q2,q3}. In fact, large scale crust cracking in neutron stars and their relevance in the occurrence of glitches and bursts of x-rays and gamma rays have been considered in detail  in \cite{q4}.

The notion of complexity of a given fluid distribution involves two different problems. On  the one hand the complexity of the structure of the fluid, which is described by the complexity factor. On the other hand, when dealing with systems in the dynamic regime we still need to describe the complexity of its  pattern of evolution. The complexity factor  is a scalar function (for non--spherical distributions  complexity may be described by more than one scalar \cite{axc}) intended to measure the degree of complexity of the structure of a self--gravitating  fluid distribution. This concept  has received a great deal of attention in recent years. The origin of such an interest being the conviction that a variable measuring complexity should  be suitable  to  describe essential aspects of the system. 
Regarding the pattern of evolution, we need to know what is the simplest mode of evolution. In \cite{ps2, epjc} two different patterns of evolutions were considered as the ``simplest'' ones, namely: the homologous ($H$) and the quasi--homologous regime ($QH$).

Finally,  we know that  dissipation due to the emission of massless particles (photons 
and/or neutrinos) is a characteristic process in the evolution of  massive stars. 

In fact, it seems that the only plausible mechanism to carry 
away the bulk of the binding energy of the collapsing star, leading 
to a neutron star or black hole is neutrino emission \cite{1}. 

We shall describe dissipation in the  diffusion approximation, which applies whenever the energy flux of radiation (as that of 
thermal conduction) is proportional to the gradient of temperature. This 
assumption is in general very sensible, since the mean free path of 
particles responsible for the propagation of energy in stellar 
interiors is in general very small as compared with the typical 
length of the object.

In many other circumstances, the mean free path of particles transporting energy may be large enough as to justify the  free streaming approximation, however  it is a simple matter to include this regime of radiative transport, just by redefining the energy--density and the radial pressure (see below). 

The purpose of this work is threefold. On the one hand, since  the treatment of cracking, so far, has been handled using non--comoving coordinates, we want to present an approach in comoving coordinates, which many authors consider more suitable for treating numerical problems. In our case, the motivation behind this endeavor is based on the fact  that  the concept of complexity for dynamical systems has been developed using comoving coordinates \cite{ps2,epjc}. It is worth mentioning that although the general idea underlying the concept of cracking remains the same in both frames, it is expressed through variables which are not exactly equivalent.

Next, we want to extend the concept of cracking, as defined in \cite{herrera}, to the dissipative case. More specifically we want to find out what might be the  role of dissipative processes in the occurrence of cracking. 

Finally,  we want to relate the concept of cracking to the concept of complexity (see \cite{ec, zhh} in relation with this issue). In particular  we want to know what constraints on the occurrence of cracking may appear from the vanishing complexity factor condition and/or from conditions on the complexity of the pattern of evolution  when leaving the equilibrium. The motivation for such an endeavor becomes intelligible if we notice that the appearance of cracking in a given self--gravitating fluid distribution implies an increasing of  complexity in the structure of the fluid, as compared with the situation when cracking is absent.

All the obtained results concerning the three issues  mentioned above, are discussed in detail in the last section.

Let us start by introducing the notation, conventions and all the required equations.
\section{Basic equations and variables}
 
In this section we shall deploy the relevant equations and variables for describing a  time dependent, dissipative, spherically symmetric self--gravitating locally anisotropic fluid. To avoid repeating calculations, the procedure to obtain some equations is referred to previous works.

\subsection{Einstein equations, physical variables, kinematical variables}

We consider spherically symmetric distributions of collapsing
fluid, which for the sake of completeness we assume to be locally anisotropic,
bounded by a spherical surface $\Sigma$, and undergoing dissipation in the
form of heat flow (diffusion approximation).

The reason to consider anisotropic fluids is well justified since  local anisotropy of pressure may be caused by a large variety of physical phenomena, of the kind we expect in compact objects \cite{rep}. More so, as it has been recently shown \cite{ps},  physical processes  expected to play a relevant role in stellar evolution (e.g. dissipation)   will always tend to produce pressure anisotropy, even if the system is initially assumed to be isotropic.  Since any equilibrium configuration is the final stage of a dynamic  regime,  there is no reason to think that the acquired anisotropy during this dynamic process   would disappear in the final equilibrium state, and therefore the resulting configuration, even if  initially  had isotropic pressure, should in principle exhibit pressure anisotropy.

Choosing comoving coordinates, the general
interior metric can be written
\begin{equation}
ds^2=-A^2dt^2+B^2dr^2+R^2(d\theta^2+\sin^2\theta d\phi^2),
\label{1}
\end{equation}
where $A$, $B$ and $R$ are functions of $t$ and $r$ and are assumed
positive. We number the coordinates $x^0=t$, $x^1=r$, $x^2=\theta$
and $x^3=\phi$. Observe that $A$ and $B$ are dimensionless, whereas $R$ has the same dimension as $r$.

The energy--momentum tensor $T_{\alpha\beta}$ of the fluid distribution
has the form
\begin{eqnarray}
T_{\alpha\beta}&=&(\mu +
P_{\perp})V_{\alpha}V_{\beta}+P_{\perp}g_{\alpha\beta}+(P_r-P_{\perp})\chi_{
\alpha}\chi_{\beta}\nonumber \\&+&q_{\alpha}V_{\beta}+V_{\alpha}q_{\beta}
, \label{3}
\end{eqnarray}
where $\mu$ is the mass-energy
density, $P_r$ the radial pressure,
$P_{\perp}$ the tangential pressure, $q^{\alpha}$ the heat flux, $V^{\alpha}$ the four-velocity of the fluid,
and $\chi^{\alpha}$ a unit four-vector along the radial direction. These quantities
satisfy
\begin{eqnarray}
V^{\alpha}V_{\alpha}=-1, \;\; V^{\alpha}q_{\alpha}=0, \;\; \chi^{\alpha}\chi_{\alpha}=1,\;\;
\chi^{\alpha}V_{\alpha}=0.
\end{eqnarray}

Or, in the equivalent (canonical) form
\begin{equation}
T_{\alpha \beta} = {\mu} V_\alpha V_\beta + P h_{\alpha \beta} + \Pi_{\alpha \beta} +
q \left(V_\alpha \chi_\beta + \chi_\alpha V_\beta\right), \label{Tab}
\end{equation}
with
$$ P=\frac{P_{r}+2P_{\bot}}{3}, \qquad h_{\alpha \beta}=g_{\alpha \beta}+V_\alpha V_\beta,$$

$$\Pi_{\alpha \beta}=\Pi\left(\chi_\alpha \chi_\beta - \frac{1}{3} h_{\alpha \beta}\right), \qquad \Pi=P_{r}-P_{\bot},$$ where $q$ is a function of $t$ and $r$.

Since we are considering comoving observers, we have
\begin{eqnarray}
V^{\alpha}&=&A^{-1}\delta_0^{\alpha}, \;\;
q^{\alpha}=qB^{-1}\delta^{\alpha}_1, \;\;
\chi^{\alpha}=B^{-1}\delta^{\alpha}_1. 
\end{eqnarray}

It is worth noticing that  both bulk and shear viscosity could be easily introduced to the system through a redefinition of  the radial and tangential pressures, $P_r$ and
$P_{\perp}$. Also,   dissipation in the free streaming approximation could be introduced by redefining  $\mu,  P_r$ and $q$. 

The Einstein equations for (\ref{1}) and (\ref{Tab}), are explicitly written  in Appendix A.

The acceleration $a_{\alpha}$ and the expansion $\Theta$ of the fluid are
given by
\begin{equation}
a_{\alpha}=V_{\alpha ;\beta}V^{\beta}, \;\;
\Theta={V^{\alpha}}_{;\alpha},\label{4b}
\end{equation}
and its  shear $\sigma_{\alpha\beta}$ by
\begin{equation}
\sigma_{\alpha\beta}=V_{(\alpha
;\beta)}+a_{(\alpha}V_{\beta)}-\frac{1}{3}\Theta h_{\alpha\beta},
\label{4a}
\end{equation}
from which we easily obtain 

\begin{equation}
a_1=\frac{A^{\prime}}{A}, \;\; a=\sqrt{a^{\alpha}a_{\alpha}}=\frac{A^{\prime}}{AB}, \label{5c}
\end{equation}

\begin{equation}
\Theta=\frac{1}{A}\left(\frac{\dot{B}}{B}+2\frac{\dot{R}}{R}\right),
\label{5c1}
\end{equation}

\begin{equation}
\sigma_{11}=\frac{2}{3}B^2\sigma, \;\;
\sigma_{22}=\frac{\sigma_{33}}{\sin^2\theta}=-\frac{1}{3}R^2\sigma,
 \label{5a}
\end{equation}
where
\begin{equation}
\sigma^{\alpha\beta}\sigma_{\alpha\beta}=\frac{2}{3}\sigma^2,
\label{5b}
\end{equation}
with
\begin{equation}
\sigma=\frac{1}{A}\left(\frac{\dot{B}}{B}-\frac{\dot{R}}{R}\right),\label{5b1}
\end{equation}
where the  prime stands for $r$
differentiation and the dot stands for differentiation with respect to $t$.

Next, the mass function $m(t,r)$ introduced by Misner and Sharp
\cite{Misner}  reads
\begin{equation}
m=\frac{R^3}{2}{R_{23}}^{23}
=\frac{R}{2}\left[\left(\frac{\dot R}{A}\right)^2-\left(\frac{R^{\prime}}{B}\right)^2+1\right],
 \label{17masa}
\end{equation}
and introducing the proper time derivative $D_T$
given by
\begin{equation}
D_T=\frac{1}{A}\frac{\partial}{\partial t}, \label{16}
\end{equation}
we can define the velocity $U$ of the collapsing
fluid  as the variation of the areal radius with respect to proper time, i.e.
\begin{equation}
U=D_TR, \label{19}
\end{equation}
where $R$ defines the areal radius of a spherical surface inside the fluid distribution (as
measured from its area).

Then (\ref{17masa}) can be rewritten as
\begin{equation}
E \equiv \frac{R^{\prime}}{B}=\left(1+U^2-\frac{2m}{R}\right)^{1/2}.
\label{20x}
\end{equation}
Using (\ref{20x}) we can express (\ref{17a}) as
\begin{equation}
4\pi q=E\left[\frac{1}{3}D_R(\Theta-\sigma)
-\frac{\sigma}{R}\right],\label{21a}
\end{equation}
where   $D_R$ denotes the proper radial derivative,
\begin{equation}
D_R=\frac{1}{R^{\prime}}\frac{\partial}{\partial r}.\label{23a}
\end{equation}
Using (\ref{12})-(\ref{14}) with  (\ref{23a}) we obtain from
(\ref{17masa})

\begin{eqnarray}
D_Rm=4\pi\left(\mu+q\frac{U}{E}\right)R^2,
\label{27Dr}
\end{eqnarray}
which implies
\begin{equation}
m=4\pi\int^{r}_{0}\left( \mu +q\frac{U}{E}\right)R^2R^\prime dr, \label{27intcopy}
\end{equation}
satisfying the regular condition  $m(t,0)=0$.

Integrating (\ref{27intcopy}) we find
\begin{equation}
\frac{3m}{R^3} = 4\pi {\mu} - \frac{4\pi}{R^3} \int^r_0{R^3\left(D_R{ \mu}-3 q \frac{U}{RE}\right) R^\prime dr}.
\label{3m/R3}
\end{equation}
\\

\subsection{The complexity factor and the Weyl tensor}
As we have already mentioned, in the dynamic case the definition of a quantity measuring the complexity of the system poses two  additional problems with respect to the static case. 

On the one hand, the definition of the complexity of the structure of the fluid, which in this case also involves  dissipative variables, and on the other hand the problem of defining    the complexity of the pattern of evolution of the system.

 For the static fluid distribution it was assumed in \cite{ps1} that  the scalar function $Y_{TF}$, appearing in the orthogonal splitting of the Riemann tensor, and named complexity factor, is an appropriate measure of the complexity of the fluid, and therefore was identified as  the complexity factor. 
 
 As in \cite{ps2}, we shall assume in the dynamic case  that $Y_{TF}$ still measures the complexity of the system, in what corresponds to the structure of the object.
 
 In order to provide the necessary mathematical expressions for defining $Y_{TF}$, let us start by finding the expression for the Weyl tensor.
 
In  the spherically symmetric case the Weyl tensor  ($C^{\rho}_{\alpha
\beta
\mu}$) is   defined by its ``electric'' part 
 $E_{\gamma \nu }$ alone, since its  ``magnetic'' part 
vanishes, with
  \begin{equation}
E_{\alpha \beta} = C_{\alpha \mu \beta \nu} V^\mu V^\nu,
\label{elec}
\end{equation}
where the electric part of the Weyl tensor  may also be written as
\begin{equation}
E_{\alpha \beta}={\cal E} (\chi_\alpha \chi_\beta-\frac{1}{3}h_{\alpha \beta}),
\label{52}
\end{equation}
with
\begin{widetext}
\begin{eqnarray}
{\cal E}&=& \frac{1}{2 A^2}\left[\frac{\ddot R}{R} - \frac{\ddot B}{B} - \left(\frac{\dot R}{R} - \frac{\dot B}{B}\right)\left(\frac{\dot A}{A} + \frac{\dot R}{R}\right)\right]+ \frac{1}{2 B^2} \left[\frac{A^{\prime\prime}}{A} - \frac{R^{\prime\prime}}{R} + \left(\frac{B^{\prime}}{B} + \frac{R^{\prime}}{R}\right)\left(\frac{R^{\prime}}{R}-\frac{A^{\prime}}{A}\right)\right] \nonumber \\&- &\frac{1}{2 R^2}.
\label{Ea}
\end{eqnarray}
\end{widetext}

Then, it can be shown that (see \cite{ps2} for details) 
\begin{eqnarray}
Y_{TF}={\cal E}-4\pi \Pi .\label{EY}
\label{EX}
\end{eqnarray}
  
Next, using  (\ref{12}), (\ref{14}), and  (\ref{15}) with (\ref{17masa}) and (\ref{Ea}) we obtain
\begin{equation}
\frac{3m}{R^3}=4\pi \left({\mu}-\Pi \right) - \cal{E},
\label{mE}
\end{equation}
which combined with (\ref{3m/R3})  and (\ref{EY}) produces

\begin{equation}
Y_{TF}= -8\pi\Pi +\frac{4\pi}{R^3}\int^r_0{R^3\left(D_R {\mu}-3{q}\frac{U}{RE}\right)R^\prime d r}.
\label{Y}
\end{equation}

Thus the scalar $Y_{TF}$ may be expressed through the Weyl tensor and the anisotropy of pressure  or in terms of the anisotropy of pressure, the mass-energy
density inhomogeneity and  the dissipative variables. 

Another useful expression for $Y_{TF}$ may be obtained (see  \cite{ss} for details), which reads

\begin{equation}
Y_{TF}\equiv {\cal E} - 4\pi \Pi  = \frac{a^\prime}{B} - \frac{\dot{\sigma}}{A} +a^2 - \frac{\sigma^2}{3} - \frac{2}{3} \Theta \sigma - a \frac{R'}{RB}\;.
\label{ss}
\end{equation}

Once the complexity factor for the structure of the fluid distribution has been established, it remains to elucidate what is the simplest pattern of evolution.

From  the integration of (\ref{13}) one obtains

\begin{equation}
U=\frac{U_{\Sigma}}{R_{\Sigma}}R-R\int^{r_{\Sigma}}_r\left(\frac{4\pi}{E} q+\frac{\sigma}{ R}\right)R^{\prime}dr,
\label{nm}
\end{equation}
where $r=r_\Sigma=constant$ is the equation of the boundary surface of the fluid distribution, and subscript $\Sigma$ means that the quantity is evaluated on the boundary surface.

If the integral in the above equation vanishes  we have that $U\sim R$, which is a reminiscence  of the homologous evolution in Newtonian hydrodynamics. This may occur if the fluid is shear--free and non dissipative, or if the two terms in the integral cancel each other.

 In the past,  two regimes of evolution have been considered as candidates to describe the simplest mode of evolution. One is the relativistic version of homologous evolution ($H$)  characterized by the conditions (see \cite{ps2} for details)

\begin{equation}
 U= \tilde a(t) R.
 \label{ven6}
 \end{equation}
 where $\tilde a\equiv\frac{U_\Sigma}{R_\Sigma}$, and 

\begin{equation}
\frac{R_I}{R_{II}}=\mbox{constant},
\label{vena}
\end{equation}
where $R_I$ and $R_{II}$ denote the areal radii of two concentric shells ($I,II$) described by $r=r_I={\rm constant}$, and $r=r_{II}={\rm constant}$, respectively. In Newtonian hydrodynamics a linear dependence of radial velocity on the radial distance, implies a condition similar to (\ref{vena}). However in the relativistic regime, (\ref{ven6}) does not imply (\ref{vena}), except in the geodesic case.

However the $H$ condition may be  too stringent,  ruling out  many interesting scenarios from the astrophysical point of view and therefore, another possible (less restrictive) mode of evolution which also could be used to describe the simplest mode of evolution, and which we call quasi--homologous ($QH$), has been proposed \cite{epjc}.
 
In this case the fluid  satisfies condition (\ref{ven6}), but not (\ref{vena}).

It follows from (\ref{nm})  that condition (\ref{ven6}), implies
\begin{equation}
\frac{4\pi}{R^\prime}B  q+\frac{\sigma}{ R}=0,
\label{ch1}
\end{equation}
which is the only condition imposed in the $QH$ regime.

To summarize, the $H$ condition   implies (\ref{vena}), and  (\ref{ch1}),  whereas the $QH$ regime only demands (\ref{ch1}).

If we impose the $H$ condition,  then it can be shown that  (see \cite{ps1} for details) 
\begin{equation}
\frac{ \ddot R}{R}-\frac{ \ddot B}{B}=Y_{TF}.
 \label{33bisa}
\end{equation}

If we further assume  the fluid to be non--dissipative, recalling that in this case the $H$ condition implies the vanishing of the shear, we obtain (see \cite{ps1} for details)
\begin{equation}
\frac{ \ddot R}{R}-\frac{ \ddot B}{B}=0\quad \Rightarrow Y_{TF}=0.
 \label{33bis}
\end{equation}

In other words, in this particular case, the $H$ condition already implies the vanishing complexity factor condition.

More so, for the non--dissipative case, the $H$ condition not only implies  $Y_{TF}=0$, but  also  implies that the fluid is shear--free, geodesic (non--dissipative dust)   with homogeneous mass-energy density and vanishing Weyl tensor, representing   the simplest conceivable configuration (Friedman--Robertson--Walker) (see \cite{gs,sc}).

Based on all the precedent comments, it seems reasonable to  consider  the $H$ condition  as a good candidate to describe the simplest  mode of evolution.

In the dissipative case, we may obtain from (\ref{5b1}) and (\ref{33bis}), 

\begin{equation}
Y_{TF}\frac{R^\prime}{R}=4\pi Bq\left(\frac{\dot{q}}{q}+2\frac{\dot B}{B}+\frac{\dot R}{R}\right).
\label{38bis}
\end{equation}

If we assume $Y_{TF}=0$, then  we obtain
\begin{equation}
 q=\frac{f(r)}{B^2R},
\label{39bis}
\end{equation}
where $f$ is an arbitrary integration function.

Taking the time derivative of the above equation and using (\ref{5c1}) and (\ref{5b1}), it follows at once 

\begin{equation}
 \dot q=-q(\Theta+\sigma).
\label{39bisb}
\end{equation}

In the above we have assumed the $H$ condition in order to describe the simplest mode of evolution, however as indicated before, such a condition may be  too restrictive, and it could be wise to consider less stringent conditions. That's why we shall also consider  the $QH$  condition (\ref{ch1}), as an alternative to describe the simplest mode of evolution.

In the dissipative case we need to provide a transport equation to describe the evolution and  distribution of temperature.
Assuming a causal dissipative theory (e.g. the Israel--Stewart theory \cite{Is1,Is2,Is3}), the transport equation for the heat flux reads
\begin{equation}
\tau h^{\alpha \beta}V^\gamma q_{\beta;\gamma}+q^\alpha=-\kappa h^{\alpha \beta}\left(T_{,\beta}+Ta_\beta\right)-\frac{1}{2}\kappa T^2 \left(\frac{\tau V^\beta}{\kappa T^2}\right)_{;\beta} q^\alpha,
\label{tre}
\end{equation}
where $\kappa$ denotes the thermal conductivity, and $T$ and $\tau$ denote temperature and relaxation time, respectively. 

In the non--relativistic regime the above equation leads to the Cattaneo-type equation \cite{cat}

\begin{equation}
\quad \tau \frac{\partial \vec q}{\partial t} + \vec q =
- \kappa \vec \nabla T\quad,
\end{equation}
which in turn produces a hyperbolic equation for the temperature (the telegraph equation) \cite{jp}

\begin{equation}\label{te}
 \frac{\kappa}{\tau \gamma} \nabla^2 T =
\frac{\partial^2 T}{\partial t^2 } + \frac{1}{\tau} \; \frac{\partial
T}{\partial t},
\end{equation}
where $\gamma$ denotes  the heat capacity per volume unit.

In the spherically symmetric case under consideration, the transport equation has only one independent component, which may be obtained from (\ref{tre}) by contracting with the unit spacelike vector $\chi^\alpha$, producing
\begin{equation}
\tau V^\alpha  q_{,\alpha}+q=-\kappa \left(\chi^\alpha T_{,\alpha}+T a\right)-\frac{1}{2}\kappa T^2\left(\frac{\tau V^\alpha}{\kappa T^2}\right)_{;\alpha} q.
\label{5}
\end{equation}

\section{Setup of the problem}
We consider a fluid distribution which is initially (say at $t=0$) in equilibrium, and then at $t=0$, due to perturbations, it  is forced to leave the equilibrium state. We shall evaluate the system in the time interval $(0, \tilde t)$, such that $\tilde  t$ is smaller than the hydrostatic time and the thermal relaxation time. Therefore in that time interval, we have
\begin{eqnarray}
\dot R=\dot B=U=\Theta=\sigma=q=0, \nonumber \\ \ddot R\neq 0 ,\quad  D_TU\neq 0, \quad D_Tq\neq 0.
\label{1sp}
\end{eqnarray}
\begin{equation}
B=B_0, R=R_0,  A=A_0,
\label{2s}
\end{equation}
where the subscript $0$ indicates the value  of the quantity in the equilibrium, and  (\ref{14})  and (\ref{28})  have been used.

To summarize: at the timescale considered here, the metric variables conserve the same value they have before the perturbation, and their first order time derivatives vanish. Also, all kinematical variables vanish, but not so their first time derivatives

We  say that there is a cracking (overturning) at some value of $r$ (say $r=r_{cr})$ whenever $D_TU$ vanishes at  $ r=r_{cr}$, being positive (negative) for   $r>r_{cr}$ and negative (positive) for  $r<r_{cr}$.

We shall denote by $F\equiv (\mu+P_r)D_TU$ the total force applied to any fluid element, immediately  after leaving the equilibrium. Then from (\ref{3m}) (evaluated at the timescale mentioned above) we may write
\begin{widetext}
\begin{eqnarray}
F
=-\left(\mu+ P_r \right)
\left[\frac{m}{R^2}
+4\pi  P_r R\right] 
-E^2\left[D_R  P_r
+2(P_r-P_{\perp})\frac{1}{R}\right] 
-E  D_T q.
\label{3mf}
\end{eqnarray}
\end{widetext}
\subsection{Non dissipative isotropic fluid}

Let us first consider  an isotropic fluid in equilibrium, whose energy density is given by 
\begin{equation}
\mu=\xi/R^2 ,
\label{es0}
\end{equation}
where $\xi$ is a constant. 

Integrating (\ref{3mf}) for $F=P_r-P_\bot=D_Tq=0 $, we obtain for $P_r$
\begin{equation}
P_r=\frac{3\xi}{r^2}\frac{\left(1-\frac{r}{r_\Sigma} \right)}{\left(9-\frac{r}{r_\Sigma}\right)},
\label{es1}
\end{equation}
with $\xi=\frac{3}{56\pi}$, and $r=r_\Sigma$ denotes the boundary surface of the fluid distribution.

Thus our a static solution  is characterized by (\ref{es0}),   (\ref{es1}) and
\begin{equation}
R=r,\quad B=\frac{1}{\sqrt{1-8\pi \xi}}=\frac{\sqrt{7}}{2},\quad A=\sqrt{r}(9r_\Sigma - r).
\label{es2}
\end{equation}

It is worth mentioning that such a solution belongs to the type $VI$ Tolman class \cite{Tolm}, whose equation of state for large values of $\mu$ approaches that for a highly  compressed Fermi gas. Since it is singular at $r=0$, the center should be excluded from consideration.

Let us now remove our system from equilibrium by perturbing the parameter $\xi$, assuming
\begin{equation}
\xi=\frac{3}{56\pi}+\epsilon,
\label{Ke}
\end{equation}
where $\vert\epsilon\vert<<1$.
It is important to stress that such a perturbation concerns only the physical variables, the metric functions remaining the same as for the static situation.

Then feeding (\ref{Ke}) back into (\ref{3mf}) we obtain 
\begin{equation}
D_T U=-28 \pi \epsilon \frac{(3-\frac{r}{r_\Sigma})}{r(9-\frac{r}{r_\Sigma})},
\label{UpKe}
\end{equation}
where we have neglected terms of order $\vert\epsilon^2\vert$ and higher, and we have assumed that the system abandons the equilibrium without dissipation.

As is apparent from (\ref{UpKe}), $D_T U$ does not change its sign in the whole interval $(0,r_\Sigma)$, implying that the system does not  endure a cracking.
\subsection{Non dissipative anisotropic fluid}
Let us now consider the anisotropic case. For doing that we shall assume  for the anisotropic factor the expression
\begin{equation}
P_r-P_\bot=\frac{\chi}{r^2},
\label{anis1}
\end{equation}
where $\chi=\frac{\xi}{4}$ and as before $\xi=\frac{3}{56 \pi}$.

Integrating (\ref{3mf})  we obtain for $P_r$
\begin{equation}
P_r=\frac{\xi}{r^2}\frac{\left(\sqrt{r_\Sigma} -\sqrt{r}\right)}{\left(\sqrt{r_\Sigma}-\frac{3}{7}\sqrt{r}\right)}.
\label{anis2}
\end{equation}

This solution is characterized by (\ref{es0}), (\ref{anis1}), (\ref{anis2}) and $R=r$.

Then perturbing the system by $ \xi\rightarrow \xi+\epsilon$ and $ \chi\rightarrow \frac{\xi}{4}(1+\omega)$, where $\vert\epsilon\vert$, $\vert  \omega\vert$ $<<1$, we obtain from (\ref{3m})
\begin{eqnarray}
D_T U&=&\frac{2\pi \epsilon}{3r}\frac{\left(-17 r_\Sigma +\frac{186}{7} \sqrt{r_\Sigma} \sqrt{r} -\frac{489}{49}r \right)}{\left(\sqrt{r_\Sigma}-\frac{5}{7}\sqrt{r}\right) \left(\sqrt{r_\Sigma}-\frac{3}{7}\sqrt{r}\right)}\nonumber \\&-&\frac{\omega}{7r}\frac{\left(\sqrt{r_\Sigma}-\frac{3}{7}\sqrt{r}\right)}{\left(\sqrt{r_\Sigma}-\frac{5}{7}\sqrt{r}\right)},
\label{anis3}
\end{eqnarray}
which we will rewrite as

\begin{equation}
W= \epsilon \left(-17  +\frac{186}{7} \sqrt{x}  -\frac{489}{49}x \right)- \omega \left(1-\frac{3}{7} \sqrt{x}\right)^2,
\label{anis4}
\end{equation}
where $W\equiv x \left(1-\frac{5}{7} \sqrt{x}\right) \left(1-\frac{3}{7} \sqrt{x}\right)r_\Sigma D_TU$ is non-negative in all the range $x\in[0,1]$, with $x\equiv \frac{r}{r_\Sigma}$ , and the parameters $\epsilon$ and $\omega$ have been reparametrized as $\frac{2\pi \epsilon}{3}\rightarrow \epsilon$, $\frac{\omega}{7}\rightarrow \omega$.

Some remarks are in order at this point:
\begin{itemize}
\item There is no cracking (overturning) if $\epsilon=0$ or $\omega=0$.
\item There is no cracking (overturning) if $\epsilon=\omega$.
\item There is no cracking (overturning)  if $\epsilon$ and $\omega$ have the same sign.
\end{itemize}

The occurrence of cracking may be easily illustrated in this case by assuming $\omega=-\delta \epsilon$, where $\delta$ is a positive real number. In this case (\ref{anis4}) becomes

\begin{equation}
W= \epsilon \left[\left(-17  +\frac{186}{7} \sqrt{x}  -\frac{489}{49}x \right)+\delta \left(1-\frac{3}{7} \sqrt{x}\right)^2\right].
\label{anis5}
\end{equation}

Figure 1 depicts function $W$ for six different values of $\delta$ in the range $(2.5,15)$. As illustrated by the figure, cracking occurs for all values of $\delta$ in the indicated range.  Also as it is apparent from this figure, greater values of $\delta$ are associated with cracking closer to the center.

Thus, while the isotropic fluid considered above leaves the equilibrium without the appearance of cracking, its anisotropic version may exhibit the occurrence  of cracking when both the radial pressure and the anisotropic factor are perturbed. Similar  conclusions were already obtained in  \cite{herrera}.
\begin{figure}[h]
\includegraphics[scale=0.7]{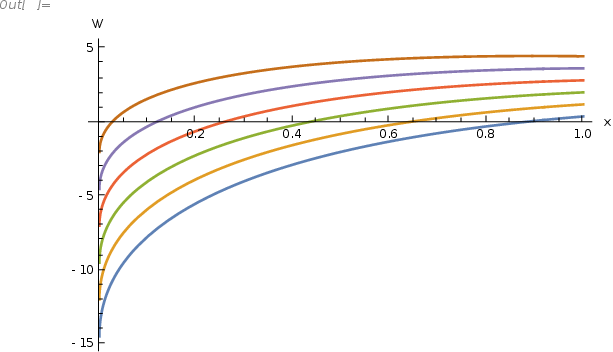}
\caption {\it  $W$ as function of $x$ for six values of $\delta$ $(2.5, 5, 7.5, 10, 12.5, 15)$. Curves from the bottom  to the top correspond to increasing values of $\delta$.}
\end{figure}

\subsection{Dissipative isotropic fluid}
We shall now turn to the case when the system leaves the equilibrium allowing the presence of dissipative processes. Since as we have just seen, pressure anisotropy (at least in the example examined above) may produce cracking, we shall consider the isotropic pressure case, in order to isolate the effects of dissipation in the possible occurrence of cracking. 

Thus evaluating the system immediately after leaving the equilibrium (``immediately'' in the sense explained above), we obtain from  the transport equation (\ref{5}) and (\ref{3m})

\begin{eqnarray}
D_T U(1-\alpha)(\mu+P_r)=&&
-(1-\alpha)(\mu+P_r)\left(\frac{m}{R^2}+4\pi P_r R\right)
\nonumber \\&-&\frac{E^2 P_r^\prime}{R^\prime}+\frac{E\kappa T^\prime}{\tau B},
\label{Updi2a}
\end{eqnarray}
where 
\begin{equation}
\alpha=\frac{\kappa T}{(\mu+P_r)\tau}.
\label{al}
\end{equation}

The last term in the above equation brings out the possible role of dissipative processes in the occurrence of cracking. However, in  order to isolate the  dissipative effects on the occurrence of cracking, we need to resort to a specific fluid distribution. For doing that we shall consider the isotropic toy model (\ref{es0}), (\ref{es1}) and (\ref{es2}).

In this case the above equation becomes
\begin{widetext}
\begin{eqnarray}
D_T U=\frac{1}{(1-\alpha)}\left[\frac{(3 r_\Sigma-r)}{r(9r_\Sigma-r)}\left(-28\pi \epsilon +\frac{6}{7}\alpha+16\pi \epsilon \alpha \right)
+\frac{E \alpha  T^\prime}{B T}\right],
\label{dis1}
\end{eqnarray}
\end{widetext}
where we have assumed $\alpha\neq 1$.

Before proceeding further we need to make some rough estimations  on the possible values of $\alpha$ defined by (\ref{al}). First of all it is worth noticing that the range of possible values of energy density, with respect to $P_r$, lie between $\mu>>P_r$ and $\mu \approx P_r$. Therefore we can neglect $P_r$ in (\ref{al}), since, at most, it would change $\alpha$ by a factor $1/2$.

So, we shall evaluate 
\begin{equation}
\alpha\approx \frac{\kappa T}{\tau \mu}.
\label{dis2}
\end{equation}

Going back from relativistic units to cgs. units we have
\begin{equation}
\kappa T=\frac{G}{c^5}(\kappa)_{c.g.s.} (T)_{c.g.s.},
\label{dis3}
\end{equation}
where $G=6.67 \times 10^{-8} g^{-1}cm^3 s^{-2}$ is the gravitational constant, $c$ is the light speed, and $(\kappa)_{c.g.s.}, (T)_{c.g.s.}$ denote the  values of $\kappa$ and $T$, expressed in $erg  \times cm^{-1}s^{-1}K^{-1}$ and $K$ (Kelvin  degrees) respectively.

Also
\begin{equation}
\tau=c (\tau)_{c.g.s.} \qquad \mu =\frac{G}{c^2}(\mu)_{c.g.s.}
\end{equation} 
where $(\tau)_{c.g.s.}$ and $(\mu)_{c.g.s.}$ denote the values of $\tau$ and $\mu$, expressed   in seconds and $g /cm^{3}$ respectively.

With all the above we may write
\begin{equation}
\frac{\kappa T}{\tau \mu}\approx 10^{-42}\frac{(\kappa)_{c.g.s.} (T)_{c.g.s.} }{(\tau)_{c.g.s.} (\mu)_{c.g.s.} }.
\label{dis5}
\end{equation}

Next, in the high frequency limit of the thermal wave, we have from the telegraph equation (\ref{te})
\begin{equation}
\tau\approx \frac{\kappa}{v^2 \gamma},
\label{dis6}
\end{equation}
where $v$ and $\gamma$ denote the speed of the thermal wave and the heat capacity per volume, respectively.

If the thermal conductivity is dominated by degenerate electrons, then we may assume for $\kappa$ \cite{fi, fib}
\begin{equation}
\kappa\approx 10^{23}\left[\mu/(10^{14}g/cm^3)\right]\left[10^8 K/T\right] erg s^{-1} cm^{-1}  K^{-1}.
\label{dis7}
\end{equation}

On the other hand, 
\begin{equation}
c_v\equiv \gamma V=\beta T,
\label{dis8}
\end{equation}
where $c_v$ is the specific heat, $V$ is the volume and for the coefficient  $\beta$ which is model dependent we assume the value proposed by Shibazaki and Lamb \cite{sl}
\begin{equation}
\beta \approx 10^{29}erg K^{-2}.
\label{dis9}
\end{equation}

Feeding back  (\ref{dis7}),  (\ref{dis8}) and  (\ref{dis9}) into (\ref{dis6}) we obtain (for densities of the order $[\mu] \approx 10^{14}$, and the radius of the degenerate core $\approx 10 Km$)
\begin{equation}
\tau\approx \frac{10^{20}}{[T^2] [v^2 ]} s,
\label{dis10}
\end{equation}
where $[\mu]$, $[T]$ and $[v]$ denote the numerical values of density, temperature and velocity of the thermal wave in $g/cm^3$, Kelvin degrees and $cm/s$ respectively.  

We shall need next to provide some possible values for the velocity of the thermal wave. 

If we take the upper limit for $v$ ($\approx 3 \times  10^{10} cm/s$), then assuming   $[T]\approx 10^2$ we obtain
\begin{equation}
\tau\approx 10^{-4}s.
\label{dis11}
\end{equation}

However, the above is probably  a too low value for the temperature, corresponding to the latest phases of the evolution of a neutron star (see \cite{sl}), and a too high value of $v$.

For a much more reasonable value of $v$ such as
\begin{equation}
v\approx 10^3 cm/s,
\label{dis12}
\end{equation}
corresponding to the value of the second sound in superfluid helium, we obtain $\tau\approx 10^{-4}s
$ for temperatures of the order of $\approx 10^9 K$, or $\tau\approx 10^{2}s$  for $T\approx 10^6 K$.

If instead we take the temperature proposed by Harwit \cite{ha} ($T\approx 10^7 K$), we obtain 
\begin{equation}
\tau\approx 1 s.
\label{dis13}
\end{equation}

To summarize, for the conditions considered above the relaxation time is in the range $(10^{-4} s, 10^2 s)$.

On the other hand, feeding back (\ref{dis7}) into (\ref{dis5}) we obtain

\begin{equation}
\frac{\kappa T}{\tau \mu}\approx 10^{-25}\frac{1}{[\tau]},
\label{dis14}
\end{equation}
or, using (\ref{dis10})
\begin{equation}
\alpha\approx\frac{\kappa T}{\tau \mu}\approx 10^{-45} [T^2][v^2].
\label{dis15}
\end{equation}

Using the above expression, we obtain for the extreme values $[T]\approx 10^{13}$ and $[v]\approx 10^9$
\begin{equation}
\alpha\approx 0.1.
\label{dis16}
\end{equation}

An alternative scenario corresponds to the early stages of a supernova during the neutronization  process. In this case the temperature may be in the range $(10^{11} K,10^{13} K)$, and the  density is about $10^{15}g/cm^3$ at the center and  $10^{13}$ on the surface. Under these conditions $\tau$ may be in the range $(10^{-6}s, 10^{-4}s)$ \cite{ma}, in which case $\alpha$ lies within the range $(10^{-4}, 10^2)$.

Although some arguments based on causality and stability conditions (see \cite{hm})  seem to prohibit values of $\alpha\geq 1$, suggesting that the value of $\alpha$, most likely, lies within  the range $(10^{-4}, <1)$, there is not a  conclusive proof about this issue. Accordingly we shall also consider the possibility of $\alpha>1$.

In order  to remove the fluid from the state of thermodynamic equilibrium we have to perturb the value of the temperature gradient corresponding to equilibrium.

Accordingly, we shall write
\begin{equation}
T=T_{eq}(1+\psi) \rightarrow T'=T'_{eq}(1+\psi), \qquad \psi \ll 1,
\label{dis19}
\end{equation}
where the subscripts $eq$ denotes the value  in equilibrium.

The condition of thermal equilibrium reads  \cite{Tolmb}
\begin{equation}
(T_{eq} A)'=0,
\label{dis20}
\end{equation}
implying
\begin{equation}
T'=-\frac{A'}{A}T_{eq}(1+\psi).
\label{cor1}
\end{equation}

Using (\ref{es2}) and (\ref{cor1}), we may write

\begin{eqnarray}
\frac{\alpha T' E}{BT}\approx  - \frac{6\alpha(3r_\Sigma-r)}{7r(9r_\Sigma-r)}(1-7\pi \epsilon),
\label{dis21}
\end{eqnarray}
where the expressions 
\begin{equation}
B\approx \frac{\sqrt{7}}{2},\qquad E\approx \frac{2}{\sqrt{7}}\left(1-7\pi \epsilon \right), \qquad \frac{T_{eq}}{T}\approx 1-\psi,
\label{dis22}
\end{equation}
have been used, and terms of order $\psi^2$ have been neglected.

Feeding back (\ref{dis21}) into  (\ref{dis1}) we obtain 

\begin{eqnarray}
D_T U=\frac{2\epsilon \pi}{(1-\alpha)}\frac{(3 r_\Sigma-r)}{r(9r_\Sigma-r)} \left[-14  +11 \alpha \right],
\label{dis24}
\end{eqnarray}
where terms of order $\epsilon^2$ have been neglected, and we recall that $\alpha$ is assumed to be different from $1$.

For cracking to occur at some point (say $r=r_{ck}$), the expression within the square bracket in (\ref{dis24}) should vanish at $r=r_{ck}$, being positive (negative) for $r>r_{ck}$ ($r<r_{ck}$), for any positive small value of $\epsilon$ (for negative values of $\epsilon$ the same argument applies for the inverse sign of the term within the square bracket).

Then for any sign of $\epsilon$,  for  values of $\alpha$ in the range $[ 10^{-4},<14/11]$,  neither cracking nor overturning will occur. 

However, if we allow the value of $\alpha$ to rise from some value within the range $[14/11 >\alpha>1]$ for $r<r_{ck}$, to the value $14/11$  for $r=r_{ck}$, increasing for $r>r_{ck}$, then we may observe a cracking produced by dissipative processes, for negative values of $\epsilon$, whereas an overturning will happen for $\epsilon>0$.

Inversely, if $\epsilon>0$, then if we allow values of $\alpha$ larger than $14/11$ for $r<r_{ck}$,  decreasing to $14/11$  for $r=r_{ck}$, cracking will also happen, whereas overturning will be observed for $\epsilon<0$. We shall discuss  these results with more detail in the last section.

\section{Cracking and complexity}
We shall next bring out the link between the restrictions imposed on the complexity of the fluid distribution and the occurrence (or not) of cracking (overturning).
As mentioned before, the analysis of the complexity of an evolving fluid implies two different (though related) issues: on the one hand the complexity of the fluid distribution, described by the scalar $Y_{TF}$ and on the other hand, the complexity of the mode of evolution. 

Regarding the complexity of the mode of evolution we shall consider the $H$ evolution described by (\ref{ven6}) and (\ref{vena}), and  the $QH$ evolution, described only by (\ref{ven6}), as the two modes of evolution to be considered as the simplest ones.

So, the question we want to answer to here is: are the vanishing complexity factor condition and the $H$ or the $QH$ evolution, compatible with the appearance of cracking (overturning)?

We shall tackle this problem by considering separately non--dissipative and dissipative fluids.
\subsection{$H$ condition, $\dot q=0$}
Let us first consider a non--dissipative fluid, and let us assume that immediately after perturbation, the system abandons the equilibrium satisfying the $H$ condition. Then we obtain from (\ref{ven6})
\begin{equation}
\dot U=\dot U_{\Sigma}\frac{R}{R_\Sigma},
\label{co1}
\end{equation}
where the fact that the system is evaluated at the timescale defined by (\ref{1sp}),   has been used.

From the above expression it is evident  that no cracking (overturning) occurs, since the sign of $\dot U$ will be the same as that of $\dot U_\Sigma$, i.e. it will be the same  for any value of $r$. On the other hand we know that the  $H$ condition in the non--dissipative case implies that $Y_{TF}=0$ (see \cite{ps2} for details).  Thus, the $H$ condition alone implies in the non--dissipative case, the vanishing of the complexity factor and prevents the occurrence of cracking.

\subsection{$QH$ condition, $\dot q=0$}
If we relax the homologous condition, assuming that  immediately after perturbation, the system abandons the equilibrium under  the quasi--homologous regime, then we obtain from (\ref{ven6})
\begin{equation}
\dot U=\dot U_{\Sigma}\frac{R}{R_\Sigma},
\label{co1b}
\end{equation}

implying again  that no cracking (overturning) occurs. However, in this case  $Y_{TF}$ does not necessarily vanish.

\subsection{$H$ condition, $\dot q\neq0$}
Let us next assume that the system abandons the equilibrium in the homologous regime, but now we allow for dissipative processes to be present. In  such  a case it follows from (\ref{38bis}), that  $Y_{TF}= 0$,  implies $\dot q =0$, i.e. dissipative processes require non--vanishing $Y_{TF}$ at the timescale under consideration.

Of course no cracking (overturning) occurs in this case since condition (\ref{co1}) is satisfied.

\subsection{$QH$ condition, $\dot q\neq0$}

If we assume instead that the system abandons the equilibrium in the $QH$ regime with  the presence of   dissipative processes,  no cracking (overturning) occurs since condition (\ref{co1}) is satisfied, but again $Y_{TF}$ does not necessarily vanish.

\subsection{$Y_{TF}=0$, $\dot q=0$, $\Pi=0$}
If we assume the vanishing complexity factor condition, without imposing any restriction on the mode of evolution, then in the isotropic non--dissipative case, we obtain that the Weyl tensor vanishes, which implies (see eq.(78) in \cite{gs})  that  the shear and its time derivatives of all orders vanish. In such  a case we obtain from (\ref{nm}), again 
\begin{equation}
\dot U=\dot U_{\Sigma}\frac{R}{R_\Sigma}.
\label{co1bc}
\end{equation}
 Thus, no cracking (overturning) is observed in this case either.

\subsection{$Y_{TF}=0$, $\dot q\neq0$, $\Pi=0$}
If we assume the vanishing complexity factor condition, without imposing any restriction on the mode of evolution, then in the isotropic dissipative case, we obtain that the Weyl tensor vanishes, however in this case we do not obtain (\ref{co1bc})  from (\ref{nm}), implying that in principle  cracking (overturning) is  allowed.

\subsection{$Y_{TF}=0$, $\dot q=0$, $a=0$}
If we assume the vanishing complexity factor condition and  the geodesic condition  $a=0$, we see from (\ref{ss}) that (at the timescale we are working with) the shear and its time derivatives of all orders vanish. In such  a case we obtain from (\ref{nm}), again 
\begin{equation}
\dot U=\dot U_{\Sigma}\frac{R}{R_\Sigma}.
\label{co1b}
\end{equation}
 Thus, no cracking (overturning) is observed in this case either.

\subsection{$Y_{TF}=0$, $\dot q\neq0$, $a=0$}
If we assume the vanishing complexity factor condition, with dissipation, then in the geodesic case $a=0$, we see from (\ref{ss}) that (at the timescale we are working with) the shear and its time derivatives of all orders vanish. However, in such  a case we do not obtain (\ref{co1b}) from (\ref{nm}), 
implying that cracking may occur in this case.

All these results as well as those obtained in the previous section, are summarized in Tables I and II.

\section{Discussion}
We started this work with three main objectives in mind. First, we wanted to  develop a general method to treat the problem of cracking in comoving coordinates. Such an approach was set up and it allowed us to tackle our second goal, namely,  to  study the influence of dissipative processes on the occurrence (or not) of cracking. Finally, using the presented approach we were able to establish the link between the occurrence (or not) of cracking and different restrictions imposed on the complexity of the fluid distribution.

Let us start the discussion, by making some general remarks on the formalism here presented:
\begin{itemize}
\item Although the basic idea underlying the concept of cracking is independent on the frame (comoving or no--comoving), the variables used to describe it, are different in both frames.
\item As for the case of non--comoving coordinates, the occurrence of cracking is described for specific solutions. However, the link between cracking and complexity was analyzed in general, without any reference to an explicit solution.
\item The influence of dissipation on the occurrence (or not) of cracking is also highly model dependent. Accordingly the example examined here is just a guide to proceed in each  specific case.
\item In order to force the system to leave the equilibrium, we shall perturb some of the parameters of the physical  variables corresponding to the solution under consideration. Then, the system is analyzed on a timescale which is smaller than the hydrostatic time and the relaxation time. In such a case we may safely assume that on this timescale, the metric functions remain the same as those before perturbation, as well as their first time derivatives, while physical variables are perturbed.
\item In relation with the comment above, particular care should be  exercised with the mass function. Indeed, as defined by (\ref{17masa}), it depends only on metric functions (and first derivatives) and then one could (wrongly) conclude that it should not be perturbed. On the other hand the mass function may also be defined through physical variables as in (\ref{27intcopy}).
This apparent contradiction is easily resolved if we remember that expressions (\ref{17masa}) and (\ref{27intcopy}) are equivalent modulo  field equations. However the metric variables after perturbation are not solutions to the field equations for the perturbed physical variables (they represent solutions for non--perturbed physical variables). Therefore, the mass function should be perturbed according to its expression (\ref{27intcopy}).
\item The concept of cracking adopted here, is based on the the definition of ``velocity'' as  given by $U$ (the areal velocity). However alternative definitions of ``velocity'' exist (see \cite{epjc})  which could be used instead of $U$, giving rise to different definitions of cracking.
\end{itemize}

Once the general setup of the problem was well defined, we proceeded to analyze first the role of  dissipative processes in the occurrence of cracking. For doing that we started  by considering  a toy model describing a static solution for an isotropic fluid  defined by (\ref{es0})--(\ref{es2}). This solution is then removed from equilibrium, by perturbing the parameter $\xi$, and we took a ``snapshot'' of the system after perturbation, on  a timescale smaller than the hydrostatic time and the thermal relaxation time. We did that assuming that no dissipative processes are allowed ensuing the perturbation. Then, it was shown that no cracking appears as a consequence of the perturbation. Next, we generalized the toy model to the case where the pressure is anisotropic, such a toy model is characterized by (\ref{es0}), (\ref{anis1}) and (\ref{anis2}). In this case we observe the appearance of cracking if both parameters $\xi$ and $\chi$ are perturbed and furthermore $\epsilon$ and $\omega$ are different and have different signs. This situation is illustrated in Figure 1.

Next, we considered the toy model described above for the isotropic fluid, but we perturbed it allowing dissipation to be present when the system abandons the equilibrium. In this case we were led to (\ref{dis24}), where the variable $\alpha$ plays a fundamental role. 

Based on some likely astrophysical scenarios, we started by making some rough estimations about the possible values of $\alpha$, as a result of which we established as a reasonable range $[10^{-4}, 10^{2}]$. However as mentioned before, some arguments based on stability seem to  rule out values of $\alpha \geq 1$ (see \cite{hm} for a discussion on this issue). Nevertheless such  arguments are not conclusive, and furthermore some numerical models with good physical behavior and $\alpha>1$ have been described in the literature (see \cite{hm1}). Accordingly we have considered also the possibility of  $\alpha>1$.

Equation (\ref{dis24}) brings out the influence of dissipation on the occurrence or not of cracking. Thus, for all values of $\alpha$ in the range $[\approx 0, <14/11]$, no cracking occurs. However, cracking (or overturning) may occur for $\alpha\approx 14/11$, for conditions summarized in Table I.

Next, we focused on our third goal, namely, to find out how the occurrence of cracking is related to restrictions imposed on the complexity of the fluid distribution. These restrictions involve the complexity factor $Y_{TF}$ and/or restrictions on the  mode in which the system leaves the equilibrium ($H$ or $QH$). 
It is important to stress here that while in the treatment of the second problem (bringing out the relevance of dissipation on the occurrence of cracking) we resorted to a specific toy model, in this third problem we obtained general results without reference to any specific solution.

The first important point to mention is that the restrictions on the mode of evolution (as the system leaves the equilibrium) appear to be  more relevant (concerning the occurrence of cracking) than the restrictions on $Y_{TF}$.

Thus, the sole  imposition of $H$ or $QH$ regime, in both, the non--dissipative and the dissipative cases, rules out the possibility of cracking. Furthermore in the former case the complexity factor vanishes. This brings out further the link between $H$ or $QH$ regime and the complexity of the mode of evolution, if we recall that the occurrence of cracking may be regarded as a factor enhancing the complexity of the system.

 We also were able to prove that the vanishing of the complexity factor alone (without any imposition on the mode of leaving the equilibrium) rules out the occurrence of cracking  in the geodesic non--dissipative case, and in the non--dissipative isotropic case. These results are summarized in Table II.

Finally, we would like to say few words about the case $\alpha=1$. As it is apparent from (\ref{Updi2a}), one of the effects of dissipative processes consists of decreasing the inertial mass density (the factor multiplying $D_TU$) and (as a consequence of the equivalent principle) the passive gravitational mass density (the factor multiplying the first square bracket on the right of (\ref{Updi2a})), by a factor $1-\alpha$. This strange effect which was discovered and discussed in \cite{hetal}, implies that the effective inertial mass density vanishes for $\alpha=1$. Until now, in spite of long discussions about this point, no definitive answer has been reached concerning  the physical meaning (if any) of this strange effect, and  the possibility of reaching the above mentioned critical value  in a real physical system.

\begin{acknowledgments}
This  work  was  partially supported by the Spanish  Ministerio de Ciencia, Innovaci\'on, under Research Project Grant PID2021-122938NB-I00 funded by MCIN/AEI/ 10.13039/501100011033 and by ``ERDF A way of making Europe''. L. H. also wishes to thank Universitat de les
 Illes Balears for financial support and hospitality. A.D.P. wishes to thank Universitat de les
 Illes Balears for its hospitality.
\end{acknowledgments}

\appendix, 
\section{Einstein equations}
 Einstein's field equations for the interior spacetime (\ref{1}) are given by
\begin{equation}
G_{\alpha\beta}=8\pi T_{\alpha\beta},
\label{2}
\end{equation}
and its non zero components
with (\ref{1}), (\ref{3}) and (\ref{5})
become
\begin{widetext}
\begin{eqnarray}
8\pi T_{00}=8\pi  \mu A^2
=\left(2\frac{\dot{B}}{B}+\frac{\dot{R}}{R}\right)\frac{\dot{R}}{R}
-\left(\frac{A}{B}\right)^2\left[2\frac{R^{\prime\prime}}{R}+\left(\frac{R^{\prime}}{R}\right)^2
-2\frac{B^{\prime}}{B}\frac{R^{\prime}}{R}-\left(\frac{B}{R}\right)^2\right],
\label{12} \\
8\pi T_{01}=-8\pi qAB
=-2\left(\frac{{\dot R}^{\prime}}{R}
-\frac{\dot B}{B}\frac{R^{\prime}}{R}-\frac{\dot
R}{R}\frac{A^{\prime}}{A}\right),
\label{13} \\
8\pi T_{11}=8\pi P_r B^2 
=-\left(\frac{B}{A}\right)^2\left[2\frac{\ddot{R}}{R}-\left(2\frac{\dot A}{A}-\frac{\dot{R}}{R}\right)
\frac{\dot R}{R}\right]
+\left(2\frac{A^{\prime}}{A}+\frac{R^{\prime}}{R}\right)\frac{R^{\prime}}{R}-\left(\frac{B}{R}\right)^2,
\label{14} \\
8\pi T_{22}=\frac{8\pi}{\sin^2\theta}T_{33}=8\pi P_{\perp}R^2
=-\left(\frac{R}{A}\right)^2\left[\frac{\ddot{B}}{B}+\frac{\ddot{R}}{R}
-\frac{\dot{A}}{A}\left(\frac{\dot{B}}{B}+\frac{\dot{R}}{R}\right)
+\frac{\dot{B}}{B}\frac{\dot{R}}{R}\right]\nonumber \\
+\left(\frac{R}{B}\right)^2\left[\frac{A^{\prime\prime}}{A}
+\frac{R^{\prime\prime}}{R}-\frac{A^{\prime}}{A}\frac{B^{\prime}}{B}
+\left(\frac{A^{\prime}}{A}-\frac{B^{\prime}}{B}\right)\frac{R^{\prime}}{R}\right].\label{15}
\end{eqnarray}
\end{widetext}
The component (\ref{13}) can be rewritten with (\ref{5c1}) and
(\ref{5b}) as
\begin{equation}
4\pi qB=\frac{1}{3}(\Theta-\sigma)^{\prime}
-\sigma\frac{R^{\prime}}{R}.\label{17a}
\end{equation}
\section{Dynamical equations}

The non trivial components of the Bianchi identities, $T^{\alpha\beta}_{;\beta}=0$, from (\ref{2}) yield
\begin{widetext}
\begin{eqnarray}
T^{\alpha\beta}_{;\beta}V_{\alpha}=-\frac{1}{A}\left[\dot { \mu}+
\left( \mu+ P_r\right)\frac{\dot B}{B}
+2\left( \mu+P_{\perp}\right)\frac{\dot R}{R}\right] 
-\frac{1}{B}\left[ q^{\prime}+2 q\frac{(AR)^{\prime}}{AR}\right]=0, \label{j4}\\
T^{\alpha\beta}_{;\beta}\chi_{\alpha}=\frac{1}{A}\left[\dot { q}
+2 q\left(\frac{\dot B}{B}+\frac{\dot R}{R}\right)\right] 
+\frac{1}{B}\left[ P_r^{\prime}
+\left(\mu+ P_r \right)\frac{A^{\prime}}{A}
+2( P_r-P_{\perp})\frac{R^{\prime}}{R}\right]=0, \label{j5}
\end{eqnarray}
\end{widetext}
or, by using (\ref{5c}), (\ref{5c1}), (\ref{16}), (\ref{23a}) and (\ref{20x}), they become, respectively,
\begin{widetext}
\begin{eqnarray}
D_T \mu+\frac{1}{3}\left(3 \mu+ P_r+2P_{\perp} \right)\Theta 
+\frac{2}{3}( P_r-P_{\perp})\sigma+ED_R q
+2 q\left(a+\frac{E}{R}\right)=0, \label{j6}\\
D_T q+\frac{2}{3} q(2\Theta+\sigma)
+ED_R  P_r 
+\left( \mu+ P_r \right)a+2(P_r-P_{\perp})\frac{E}{R}=0.
\label{j7}
\end{eqnarray}
\end{widetext}
This last equation may be further transformed as follows, the acceleration $D_TU$ of an infalling particle can
be obtained by using (\ref{5c}), (\ref{14}), (\ref{17masa})  and (\ref{20x}),
producing
\begin{equation}
D_TU=-\frac{m}{R^2}-4\pi  P_r R
+Ea, \label{28}
\end{equation}
and then, substituting $a$ from (\ref{28}) into
(\ref{j7}), we obtain
\begin{widetext}
\begin{eqnarray}
\left( \mu+ P_r\right)D_TU 
=-\left(\mu+ P_r \right)
\left[\frac{m}{R^2}
+4\pi  P_r R\right] 
-E^2\left[D_R  P_r
+2(P_r-P_{\perp})\frac{1}{R}\right] 
-E\left[D_T q+2 q\left(2\frac{U}{R}+\sigma\right)\right].
\label{3m}
\end{eqnarray}
\end{widetext}

\begin{widetext}
\begin{table}[htp]
\caption{Cracking and complexity}
\begin{center}
\begin{tabular}{ccccccccccccccccc}
\hline\hline
$ H, QH, Y_{TF}\diagdown   \dot q$ &$\quad$&
$0$ &$\quad$ &
$\neq 0$&$\quad$ &
& &$\quad$ &
 \vspace{0.3cm} \\  \hline
$H$&$\quad$&no cracking, no overturning, $Y_{TF}=0$ &$\quad$ &   no cracking, no overturning,  $Y_{TF}\neq 0$      &$\quad$ \\ \hline
$QH \rightarrow Y_{TF}\neq 0$&$\quad$&no cracking, no overturning &$\quad$ &   no cracking, no overturning      &$\quad$ \\ \hline
$Y_{TF}=0, $$\Pi$=0$$ &$\quad$ & no cracking, no overturning &$\quad$  &    cracking or  overturning  are allowed    &$\quad$ \\ \hline
$Y_{TF}=0, $a=0$$ &$\quad$ & no cracking, no overturning &$\quad$  &    cracking or  overturning  are allowed    &$\quad$ \\ \hline
\end{tabular}
\end{center}
\label{data}
\end{table}
\end{widetext}

\begin{widetext}
\begin{table}[htp]
\caption{Cracking and dissipation}
\begin{center}
\begin{tabular}{ccccccccccccccccc}
\hline\hline
$\epsilon \diagdown   \alpha$ &$\quad$&
$[\alpha \approx10^{-4}, \alpha < 14/11]$ &$\quad$ &
$[14/11 >\alpha >1, \alpha >14/11]$ increasing&$\quad$ &
$[\alpha>14/11 , 14/11>\alpha>1]$ decreasing& &$\quad$ &
 \vspace{0.3cm} \\  \hline
$\epsilon>0$&$\quad$ &no cracking, no overturning &$\quad$ &  overturning     &$\quad$ & cracking   &\\ \hline
$\epsilon<0$ &$\quad$ & no cracking, no overturning  &$\quad$ &   cracking      &$\quad$ & overturning   &$\quad$ & \\ \hline
\end{tabular}
\end{center}
\label{data}
\end{table}
\end{widetext}

\end{document}